%% file: EDTM_arxiv.tex
\documentclass[12pt,a4paper]{article} 
\usepackage[left=1.6cm, right=1.6cm, bottom=2.54cm, top=1.9cm]{geometry}
\usepackage{amsmath,amssymb,amsfonts}
\usepackage{algorithmic}
\usepackage{graphicx}
\usepackage{textcomp}
\usepackage{xcolor}
\usepackage{caption}
\usepackage{amsmath}
\usepackage{tikz, graphicx}
\usepackage{epsfig}
\usepackage{amssymb}
\usepackage{multirow}
\usepackage{float}
\usepackage{bm}
\usepackage{color}

\usepackage[lofdepth,lotdepth]{subfig}
\usepackage{ulem}
\usepackage{xfrac}
\usepackage{bigstrut}
\usepackage{multicol}
\usepackage{bigints}
\usepackage[noadjust]{cite}
\usepackage{tabularx}
\usepackage{booktabs}
\usepackage{array}
\usepackage{lettrine}
\newcolumntype{+}{>{\global\let\currentrowstyle\relax}}
\newcolumntype{^}{>{\currentrowstyle}}

\renewcommand{\arraystretch}{1.6}

\include{macros}

\begin{document}
\onecolumn
\setlength{\abovedisplayskip}{3.5pt}
\setlength{\belowdisplayskip}{3.5pt}

\pagenumbering{gobble}

\title{Modelling and design of FTJs as high reading-impedance synaptic devices}
\author{ \large{R. Fontanini, M. Massarotto, R. Specogna, F. Driussi, M. Loghi, D. Esseni}\\
	\small{DPIA, University of Udine, Via delle Scienze 206, 33100 Udine, Italy; email: david.esseni@uniud.it;}	
}
\maketitle
%
\section{Abstract}
We present an in-house modelling framework for Ferroelectric Tunnelling Junctions (FTJ), and an insightful study of the design of FTJs as synaptic devices. 
Results show that a moderately low-$\kappa$ tunnelling dielectric (e.g. SiO$_2$) can increase the read current and the current dynamic range.\\

\vspace{-2mm}
Keywords - Ferroelectric Tunnelling Junctions, Neuromorphic Computing\\
\\\textbf{© 2021 IEEE. Personal use of this material is permitted. Permission from IEEE must be obtained for all other uses, in any current or future media, including reprinting/republishing this material for advertising or promotional purposes, creating new collective works, for resale or redistribution to servers or lists, or reuse of any copyrighted component of this work in other works.}

\vspace{-1mm}
\section{Introduction}
\label{sec:Intro}
%
%

\begin{figure}[]
	\centering
	\includegraphics[width = 0.9 \columnwidth]{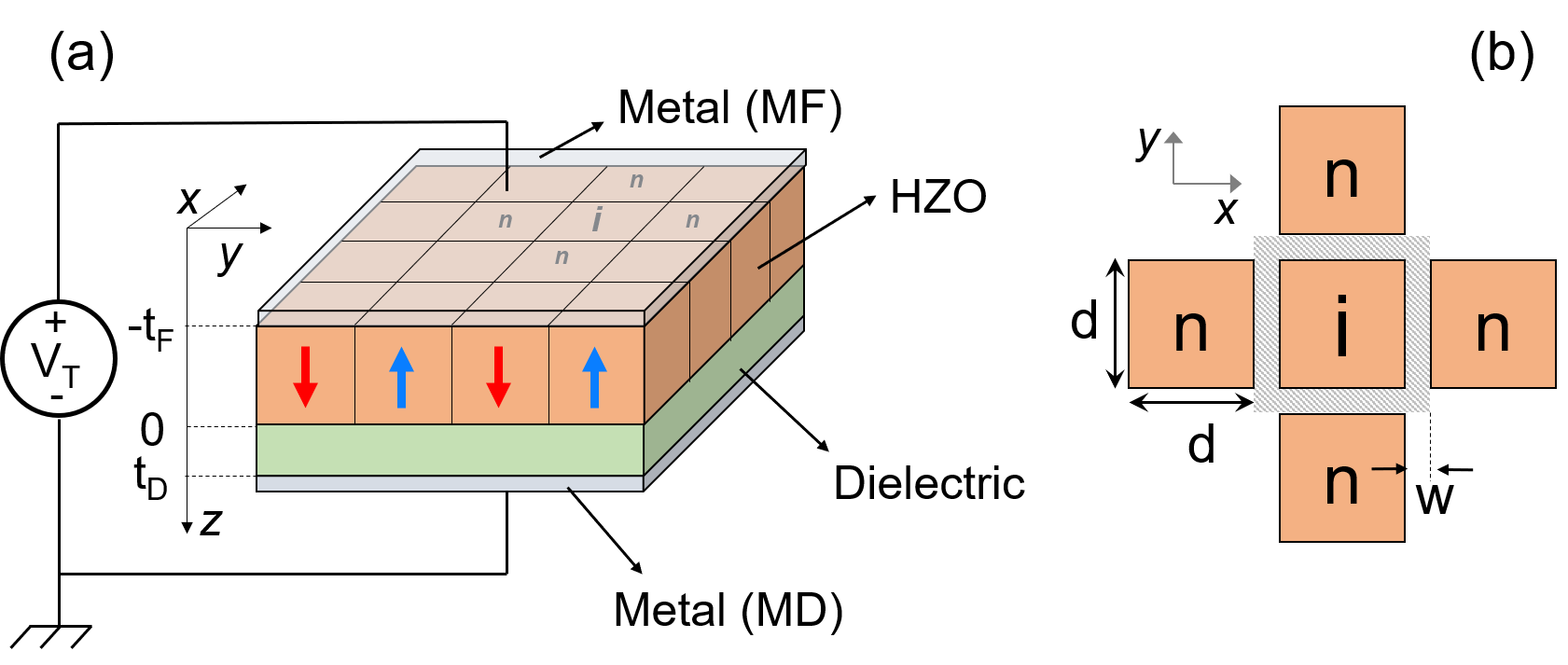}\vspace{-2mm}
	\caption{\label{Fig:FTJ_sketch} \protect \footnotesize (a) Sketch of a MFIM based FTJ, where $MF$, $MD$ are the electrodes contacting respectively the ferroelectric and the dielectric, while $t_F$ and $t_D$ are the ferroelectric and dielectric thickness. A positive ferroelectric polarization points towards the dielectric (red arrow).
		(b) Zoom on ferroelectric domains where $d$ is the side of the square domain and $w$ is the width of the domain-wall region used for the domain wall energy in LGD (Eq.1) \cite{Rollo_Nanoscale2020}. Throughout the work we used $d$$=$5 nm, $w/d =0.1$ and the domain wall coupling factor in { Eq.1} was set to $k/w =2\times10^{-3}[m^2/F]$. $V_T$ is the external bias.
		%
	}
\end{figure}

\begin{figure}[]
	\centering
	\includegraphics[width = 0.85 \columnwidth]{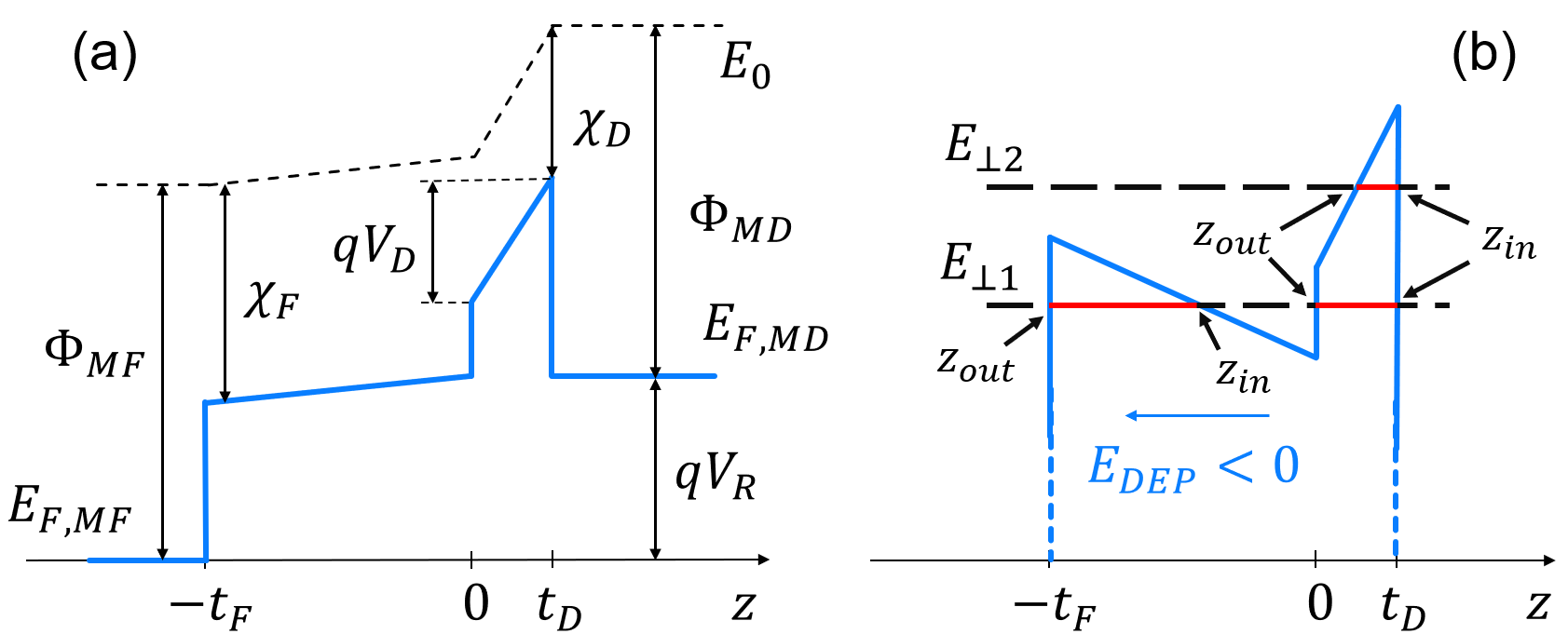}\vspace{-2mm}
	\caption{\label{Fig:Read_Retent} \protect \footnotesize Band diagram across the  MFIM stack. $E_0$, $\Phi_{MF}$, $\Phi_{MD}$ are respectively the vacuum level, and the work function of the MF and MD electrodes. $\chi_{F}$, $\chi_{D}$ are the electron affinity of the ferroelectric and dielectric, $E_{f,MD}$, $E_{f,MF}$ are the Fermi levels of  the MD and MF electrode. (a) Read condition with a read voltage $V_T$$=$$V_R$: $qV_D$ should be larger than the ferroelectric tunnelling barrier $[\Phi_{MD}$$-$$\chi_{F}]$, so that the ferroelectric conduction band profile can drop below $E_{f,MD}$; (b) Retention condition for $V_T$$=$0: the depolarization field $E_{DEP}$$\approx$$P_r\left [ \varepsilon_F \left ( C_D/C_F +1 \right )  \right ]^{-1}$ should be minimized ($P_r$ is the remnant polarization).
	}
\end{figure}

There exists a growing need for electron devices capable of non volatile, multi-level adjustments of the resistance \cite{Yu_IEEE_Proc2018}. In fact, such synaptic devices can improve by orders of magnitude the energy efficiency of neural network hardware \cite{Ambrogio_Nature2019}, for which purpose low programming energy and high reading-impedance are crucial \cite{Slesazeck_Nanotechnology2019}. Ferroelectric Tunneling Junction (FTJ) is a promising candidate as energy efficient synaptic device, and a four level operation has been experimentally reported in a Metal-Ferroelectric-Insulator-Metal (MFIM) architecture (Fig.\ref{Fig:FTJ_sketch}(a)) \cite{Max_JEDS2019}. The design of a MFIM FTJ has a delicate trade--off between the read operation, demanding a large enough dielectric voltage drop $V_D$ (i.e. $qV_D$$>$$[\Phi_{MD}$$-$$\chi_{F}]$, see Fig.\ref{Fig:Read_Retent}(a)) and thus a small dielectric capacitance $C_D$$=$$\varepsilon_D$$/$$t_D$, and the retention condition requiring instead a large $C_D$$/$$C_F$ ratio to minimize the depolarization field $E_{DEP}$$\approx$$P_r\left [ \varepsilon_F \left ( C_D/C_F +1 \right )  \right ]^{-1}$ \cite{Slesazeck_Nanotechnology2019} (Fig.\ref{Fig:Read_Retent}(b)). 
Furthermore the dielectric thickness $t_D$ and its electron affinity $\chi_D$ have a large impact on the read tunnelling current.

Due to the many material and device options, there is an urgent need for a simulation driven optimization of FTJs. However the modelling of FTJs is challenging, as it entails the ferroelectric dynamics for a three dimensional (3D) electrostatics and the tunnelling through the dielectric stack.

In this work we first present an in-house developed modelling framework for FTJs, including the 3D treatment of the ferroelectric dynamics and electrostatics, and the tunnelling transport through the MFIM stack.
The model is then validated against experiments \cite{Max_JEDS2019, Max_ESSDERC2018}, and used for an insightful design study of FTJs, to optimize the read current dynamic range.

\vspace{-1mm}
\section{Modelling approach}
\label{sec:ModelGeneral}
Our starting point is the multi-domain Landau, Ginzburg, Devonshire (LGD) model (Eq.1, $i$$=1, 2 \cdots n_D$, with $n_D$ being the number of domains) for a MFIM capacitor presented in \cite{Rollo_IEDM2018,Rollo_IEDM2019,Rollo_Nanoscale2020}.
In Eq.1 $\alpha_i$, $\beta_i$, $\gamma_i$ are the domain dependent ferroelectric anisotropy constants, $C_0$$=$$(C_D$$+$$C_F)$,
$d$ is the side of the square domain, $k$ and $w$ are the coupling constant and the inter-domain region width for the domain wall energy (Fig.\ref{Fig:FTJ_sketch}(b)), while the capacitances $C_{i,j}$ provide a three dimensional description of the depolarization energy and obey the sum rules $\sum_{j=1}^{n_D} (1/C_{i,j}) \simeq \sum_{i=1}^{n_D} (1/C_{i,j}) \simeq 1/C_{0}$ \cite{Rollo_Nanoscale2020}. 
At each time $t$ and bias $V_T(t)$, Eq.1 provides all the domain polarizations $P_i(t)$, so that the dielectric, $V_{D,i}$, and ferroelectric, $V_{F,i}$, voltage drops are uniquely given by Eqs.2 \cite{Rollo_Nanoscale2020}.
As it can be seen, the resistivity $\rho$ sets a time scale $t_{\rho}$$ = $ $\rho/ (2|\langle \alpha \rangle |)$ of the ferroelectric dynamics (where $\langle \alpha \rangle$ is the $\alpha$ value averaged over domains), so that a slow bias $V_T(t)$ compared to $t_{\rho}$ results in a quasi static behavior.

\iffalse
The tunnelling read current, \IR, is estimated as the sum of the \IRi\ in each domain, in turn given by {\bf Eq.3} resulting from a Landauer model \cite{Lunstrom_Nanotransistors}. In {\bf Eq.3} the Fermi functions  are $f_{0,MD(F)}(E)$$=$$[1+\exp(E-E_{f,MD(F)})]^{-1}$, with $E_{f,MD}$, $E_{f,MF}$ being the Fermi levels of the MD and MF electrodes defined in {\bf Fig.\ref{Fig:FTJ_sketch}(a)}.
{\bf Eq.3} assumes an effective mass approximation\footnote{
Here $m_\parallel$ corresponds to an effective mass for the density of states of the metal electrodes. In the lack of a better $m_\parallel$ determination, we used the popular assumption $m_\parallel$$\approx$$m_0$ \cite{Driussi_ME2013}. 
}
and an energy separability $E$$=$$E_{\perp}$$+$$\varepsilon(\kv)$, with the transverse energy $\varepsilon(\kv)$ being conserved in the tunnelling process \cite{Fischetti_JAP1995, Takagi_IEEE99}. For a tunnelling transmission, $T_i(E_\perp)$, independent of (\kx,\ky), the sum over (\kx,\ky) can be evaluated analytically as in {\bf Eq.4} \cite{Esseni_NanoMOS_Book2011}.

%
%
By substituting in {\bf Eq.3} one readily obtains the \IRi\ expression in {\bf Eq.5} where, for a read voltage $V_R$ applied to the MF electrode, we have $E_{f,MF}$$=$$E_{f,MD}$$-$$qV_R$. Finally the transmission $T_i(E_\perp)$ was calculated according to the WKB approximation and an effective oxide mass $m_{ox}$ ({\bf Tab.\ref{Tab:Sims_Param}}) \cite{Driussi_TED2014,Spiga_ACS2018}.
For WKB calculations the conduction band profiles $E_{CD,i}(z)$, $E_{CF,i}(z)$ in the oxide and ferroelectric are assumed to be linear and set by the $V_{D,i}$ and $V_{F,i}$ ({\bf Eqs. 2, 6}), which simplifies the determination of the tunnelling extrema $z_{in}$, $z_{out}$ ({\bf  Fig.\ref{Fig:Read_Retent}}). Depending on the band diagram, two tunnelling paths may be involved in the WKB calculation at a given $E_\perp$ (as shown in {\bf Fig.2(b)}), in which case $T_i(E_\perp)$ is obtained by the product of the two tunnelling transmissions. This approach neglects the influence on $T_i(E_\perp)$ of interference effects, which is a reasonable approximation also in virtue of the empirical calibration of some modelling parameters discussed below.
\else
The tunnelling read current, \IR, is estimated as the sum of the \IRi\ in each domain, in turn given by Eq.3 resulting from a Landauer model \cite{Lunstrom_Nanotransistors}. In Eq.3 the Fermi functions  are $f_{0,MD(F)}(E)$$=$$[1+\exp(E-E_{f,MD(F)})]^{-1}$, with $E_{f,MD}$, $E_{f,MF}$ being the Fermi levels of the electrodes as defined in Fig.\ref{Fig:FTJ_sketch}(a).
Eq.3 assumes an effective mass approximation and an energy separability $E$$=$$E_{\perp}$$+$$\varepsilon(\kv)$, with the transverse energy $\varepsilon(\kv)$ being conserved in the tunnelling process \cite{Fischetti_JAP1995}. For a tunnelling transmission, $T_i(E_\perp)$, independent of (\kx,\ky), the sum over (\kx,\ky) can be evaluated analytically as in Eq.4 \cite{Esseni_NanoMOS_Book2011},  thus obtaining the \IRi\ expression in  Eq.5 where, for a read voltage $V_R$ applied to the MF electrode, we have $E_{f,MF}$$=$$E_{f,MD}$$-$$qV_R$. Finally the transmission $T_i(E_\perp)$ is calculated accordingly with a WKB approximation and an effective oxide mass $m_{ox}$.
For WKB calculations the conduction band profiles $E_{CD,i}(z)$, $E_{CF,i}(z)$ in the oxide and ferroelectric are assumed to be linear and set by the $V_{D,i}$ and $V_{F,i}$ (Eqs.2), which simplifies the determination of the tunnelling extrema $z_{in}$, $z_{out}$ (Fig.\ref{Fig:Read_Retent}(b)). Depending on the band diagram, two tunnelling paths may be involved in the WKB calculation at a given $E_\perp$ (Fig.\ref{Fig:Read_Retent}(b)), in which case $T_i(E_\perp)$ is obtained by the product of the two tunnelling transmissions. This approach neglects the influence on $T_i(E_\perp)$ of interference effects, which is a reasonable approximation also in virtue of the empirical calibration of some modelling parameters discussed below.
\fi
%
%
\vspace{-1mm}
\section{Model validation against experiments}
\label{sec:Cdi_Negative}
\begin{figure}[]
	\centering
	\begin{minipage}[]{0.45\textwidth}
		\includegraphics[width = 1 \columnwidth]{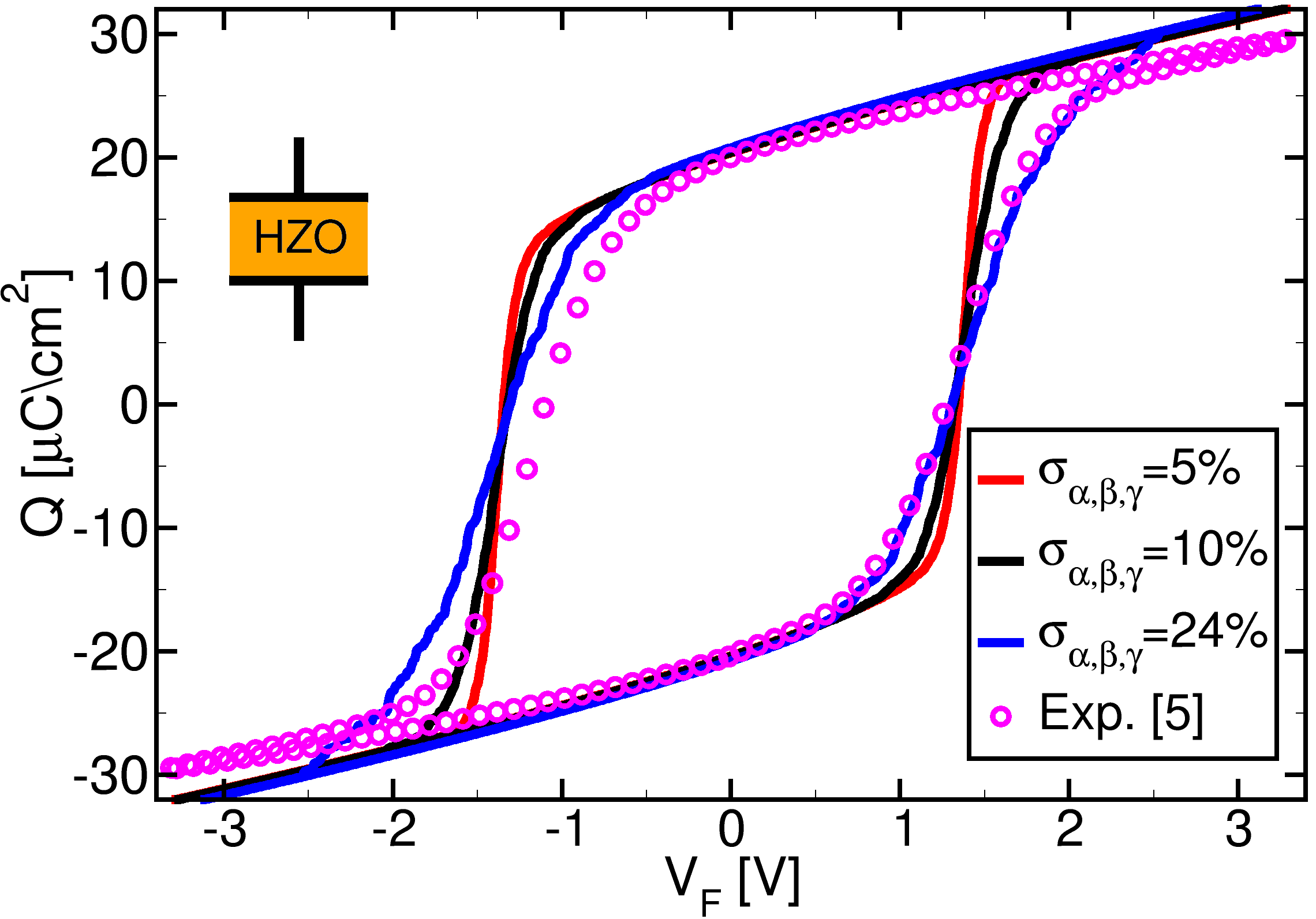}\vspace{-1mm}
		\caption{\label{Fig:PT_Vfe_Exp_Sim} \protect \footnotesize Experimental \cite{Max_ESSDERC2018} (symbols) and simulated (lines) polarization versus ferroelectric voltage characteristic of a Hf$_{0.5}$Zr$_{0.5}$O$_{2}$ layer with $t_{F}$$=$12 nm.
			The nominal values of the anisotropy constants in the LGD equations are $\alpha$$=$$-$5.8$\cdot$$10^{8}$m/F, $\beta$=$2.9\cdot$$10^{9}$m$^5$/F/C$^2$, $\gamma$$=$6.5$\cdot$$10^{10}$m$^9$/F/C$^4$, and domain to domain variations are introduced according to a normal distribution of $\alpha_i$, $\beta_i$, $\gamma_i$, where $\sigma_{\alpha,\beta,\gamma}$ denote the standard deviations normalized to the mean values. Both experiments and simulations correspond to a quasi static condition.}
	\end{minipage}\qquad
	\begin{minipage}[]{0.45\textwidth}
		\centering
		\vskip23mm
		\includegraphics[width = 1\columnwidth]{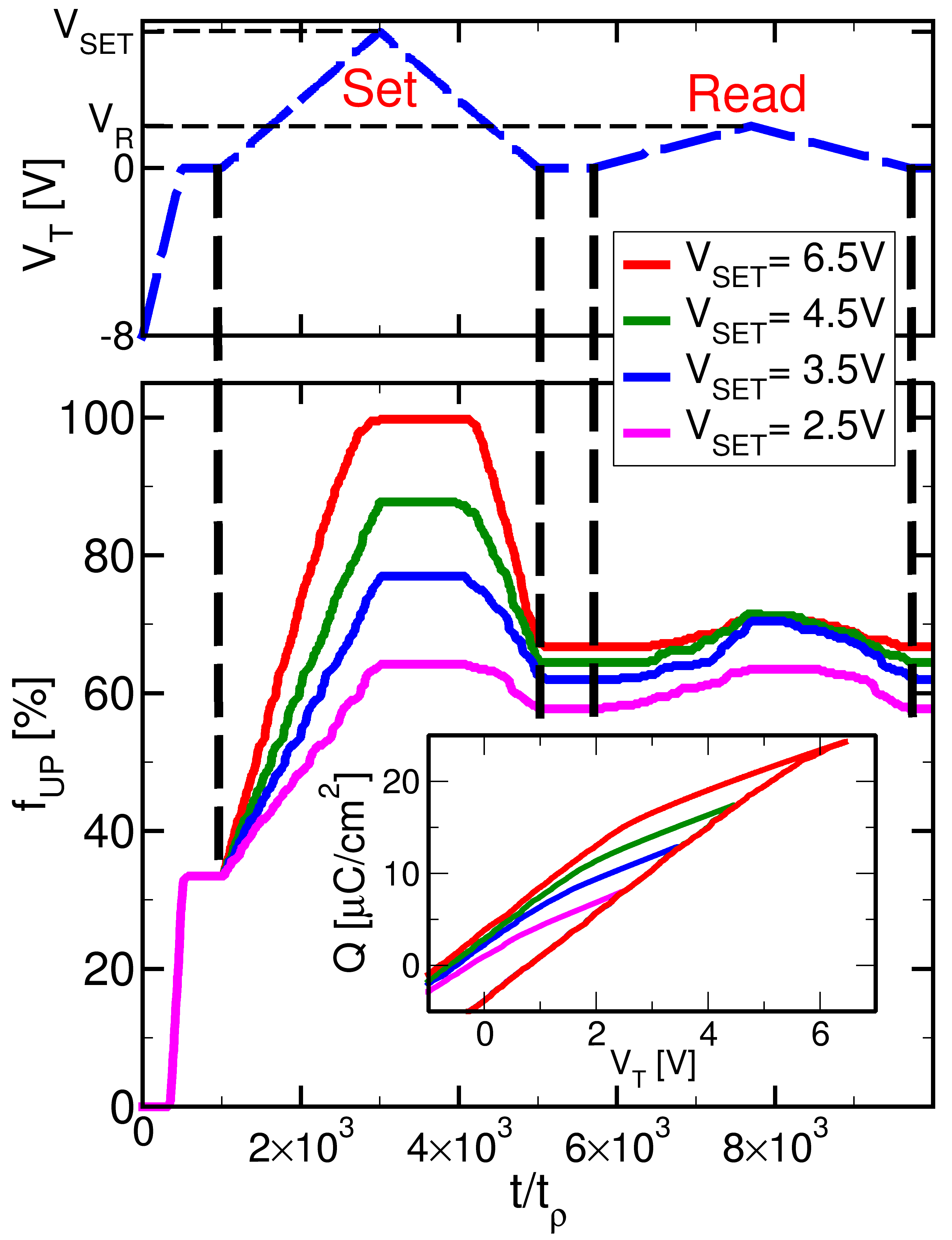}\vspace{-2mm}
		\caption{\label{Fig:Minor_Loops} \protect \footnotesize (Top) Examples of setting and reading waveforms used in the simulations of the MFIM based FTJs of \cite{Max_JEDS2019} (see parameters in Tab.\ref{Tab:Sims_Param}); read voltage is $V_R$$=$2 V. (Bottom) Corresponding fractions $f_{UP}$, defined as the percentage of the domains having a positive polarization during the set and read operation and for different $V_{SET}$. The inset shows minor loops in the $Q$ versus $V_T$ plots corresponding to different $V_{SET}$ values.}
	\end{minipage}
	\vspace{-6mm}
\end{figure}

We validated our model by comparison
with measured data for both the ferroelectric $Q$-$V_{F}$ curve (with $Q$$=$$P+\varepsilon_0 \varepsilon_{F} E_{F}$) \cite{Max_ESSDERC2018}, and the $I_R$ in corresponding FTJs \cite{Max_JEDS2019}. Fig.\ref{Fig:PT_Vfe_Exp_Sim} reports the measured $Q$-$V_{F}$ data for Hf$_{0.5}$Zr$_{0.5}$O$_{2}$ (HZO) reported
in \cite{Max_ESSDERC2018}, and the simulations from the LGD multi-domain model in Eq.1.

The agreement between simulations and experiments is good, and the match in the switching region improves by accounting for domain-to-domain variations of $\alpha_i$, $\beta_i$ and $\gamma_i$. Throughout this work we will use the nominal values for $\alpha$, $\beta$ and $\gamma$ reported in the caption of Fig.\ref{Fig:PT_Vfe_Exp_Sim}. 

We simulated a MFIM structure featuring a 12nm HZO ferroelectric, a 2nm Al$_2$O$_3$ dielectric and TiN metal electrodes \cite{Max_JEDS2019} (see Tab.\ref{Tab:Sims_Param} for material parameters). Figure \ref{Fig:Minor_Loops} (top) shows examples of the setting and reading waveforms employed for the simulations of the FTJs, that were shaped to emulate the triangular waveforms used in the experiments \cite{Max_JEDS2019}. 
As it can be seen the $V_T$ waveforms are very slow compared to $t_{\rho}$,
hence simulations correspond to a quasi static operation. 

In Fig.\ref{Fig:Minor_Loops} (bottom), different $V_{SET}$ values clearly result in different fractions, $f_{UP}$, of domains with a positive polarization, stemming from the minor loops in the $Q$ vs. $V_T$ curve shown in the inset. By inspecting the $f_{UP}$ in the set and read operation we also see that the MFIM device suffers from a quite strong depolarization effect. In fact, for a given $V_{SET}$, the $f_{UP}$ during retention (i.e. for $V_T$$=$0 V) and read (i.e. for $V_T$$=$2 V) is significantly smaller than the corresponding $f_{UP}$ reached in the set operation. This occurs because the fairly large $t_D$$=$2 nm results in a strong depolarization field (Fig.\ref{Fig:Read_Retent}(b)), producing a back-switching to $P_i$$<$$0$ of some of the domains having $P_i$$>$$0$ during the setting phase. 

Figure \ref{Fig:Itunn_Exp_Sim} finally compares simulations and experiments for the $I_R$ of FTJs at a read voltage $V_R$$=$2 V. 
\iffalse
\else
Here we notice that the $m_\parallel$ in Eq.5 corresponds to an effective mass for the density of states of the metal electrodes. In the lack of a better $m_\parallel$ determination, we used the popular assumption $m_\parallel$$\approx$$m_0$ \cite{Driussi_ME2013}. Our simulations in Fig.\ref{Fig:Itunn_Exp_Sim} can track experiments well with reasonable values of the oxide mass $m_{ox}$ (Tab.\ref{Tab:Sims_Param}) \cite{Driussi_TED2014}.
\fi

\section{Simulation based design of FTJs}
\label{sec:Cdi_Negative}

\begin{figure}[]
	\centering
	\includegraphics[width = 0.75\columnwidth]{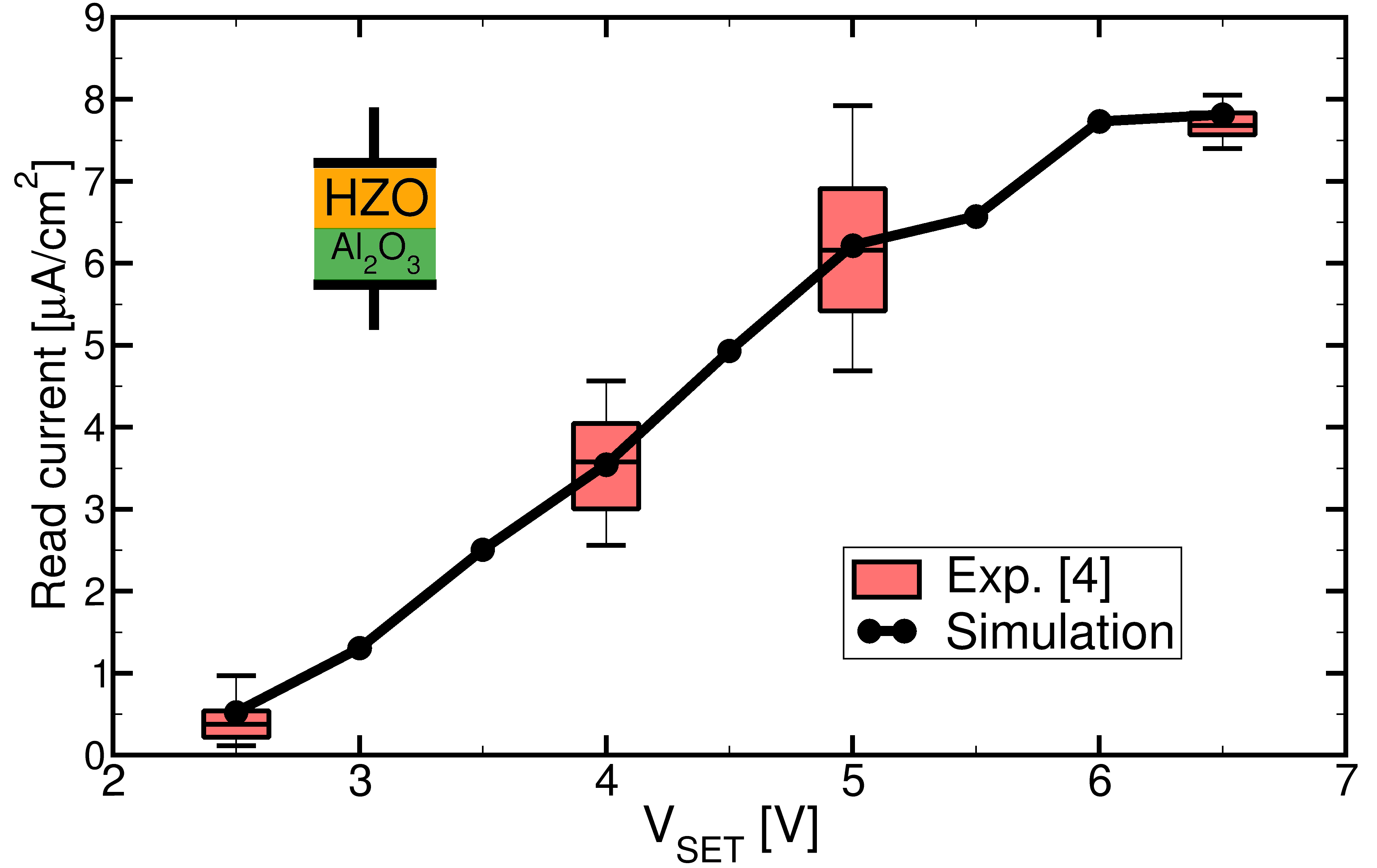}\vspace{-2mm}
	\caption{\label{Fig:Itunn_Exp_Sim} \protect \footnotesize Experiments from \cite{Max_JEDS2019} (boxes,
		device area $\approx$$3.14 \cdot 10^{-4}$ cm$^{2}$), and simulations (black solid line) for the read current at $V_R$$=$2 V of an HZO/Al$_2$O$_3$ FTJ (12nm / 2nm) and versus the set voltage $V_{SET}$. The box plots for experiments were inferred from the cycle to cycle variations reported in Fig.3(d) of \cite{Max_JEDS2019}.
	}
\end{figure}

The minimum $I_R$ value required by applications is set by the transistors leakage current and by the noise of the sense amplifier,
and for recent designs of neuromorphic processors a reasonable target is approximately 100 pA \cite{Qiao_BioCAS_2016,Sharifshazileh_ICECS2018}. The results in Fig.\ref{Fig:Itunn_Exp_Sim} show that for $V_{SET}$$=$2.5~V a device area larger than $10^{4}\ \mu m^2$ is needed to reach the $I_R$$=$100 pA. Moreover the current ratio $R_I$$=$$[I_{R,max}$$/$$I_{R,min}]$ is only about ten, that seems too small for the desired 4--bit resolution of the synaptic weights \cite{Pfeil_Front_NeuroSc_2012}. Hence the primary goals of our FTJ design exploration are the increase of $I_R$ and $R_I$.

\begin{figure}[]
	\centering
	\includegraphics[width = 0.65 \columnwidth]{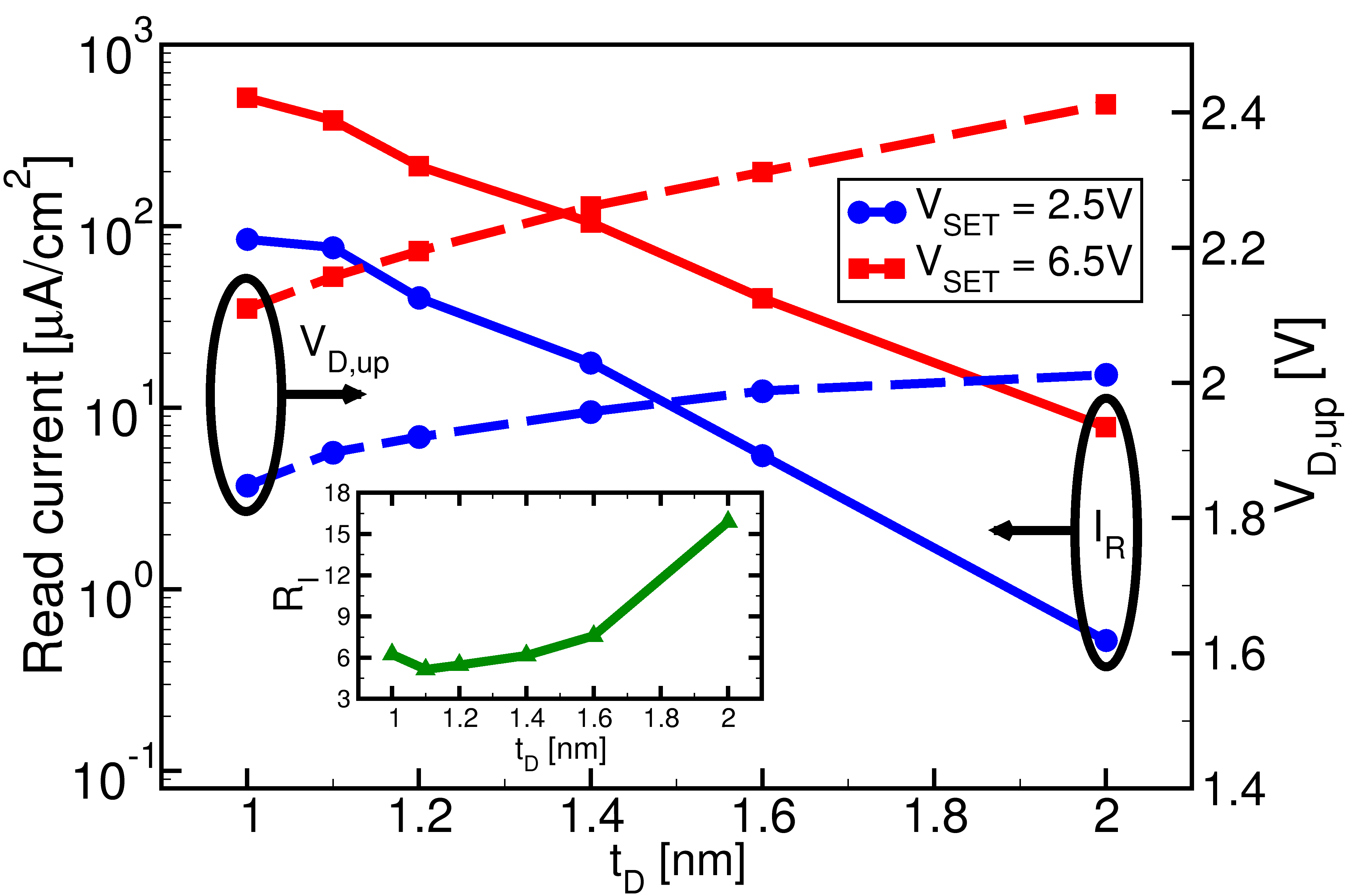}\vspace{-2mm}
	\caption{\label{Fig:IT_Tox_Al2O3} \protect \footnotesize Read current at $V_R$$=$2 V (left $y$ axis) versus the $Al_2O_3$ thickness for an HZO/Al$_2$O$_3$ FTJ ($t_F$$=$12nm) and for $V_{SET}$$=$2.5 V or 6.5 V. The average voltage drop, $V_{D,up}$, for the positive polarization domains is also reported (right $y$ axis) in read condition, and the current ratio $R_I$$=$$[I_{R,max}$$/$$I_{R,min}]$ is shown in the inset.
	}	
\end{figure}

The most obvious route to increase $I_R$ is the scaling of the dielectric thickness $t_D$, whose effects are illustrated in Fig.\ref{Fig:IT_Tox_Al2O3} for the HZO/Al$_2$O$_3$ stack. As it can be seen $I_R$ increases by thinning the Al$_2$O$_3$ layer, but $R_I$ degrades (see inset). A marked $R_I$ reduction with decreasing $t_D$ has been observed also in experiments \cite{Max_ESSDERC2018}.
Fig.\ref{Fig:IT_Tox_Al2O3} also shows that, by thinning Al$_2$O$_3$ and thus increasing $C_D$$=$$\varepsilon_d$$/$$t_D$, the average $V_{D,up}$ for the positive polarization domains decreases (right $y$-axis), and eventually $qV_{D,up}$ cannot overcome  $[\Phi_{MD}$$-$$\chi_{F}]$$\simeq$2.45 eV (see Fig.\ref{Fig:Read_Retent}(a)).

\begin{figure}[]
	\centering
	\begin{minipage}[t]{0.5\textwidth}
		\centering
		\includegraphics[width = 1 \textwidth]{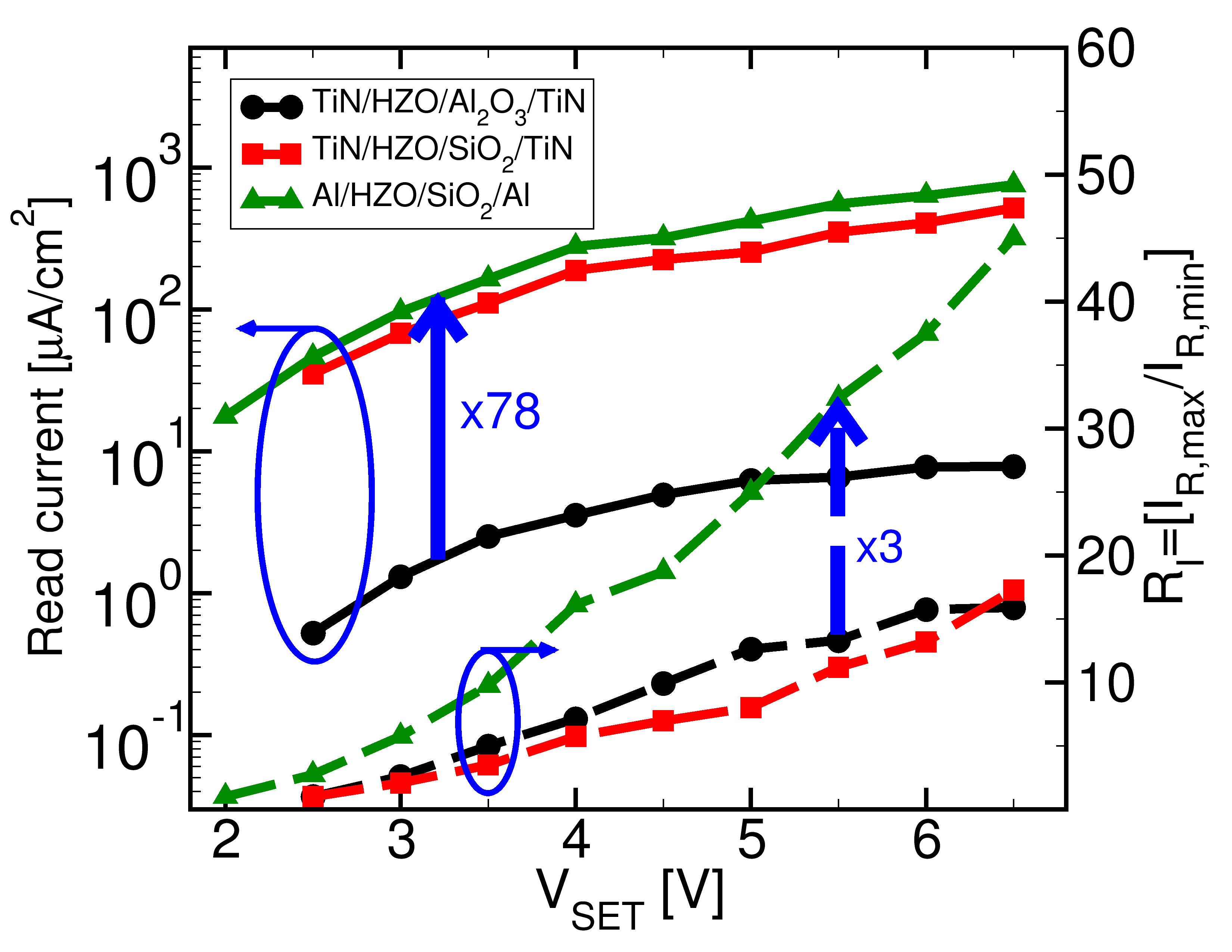}			
	\end{minipage}\hfill
	\begin{minipage}[t]{0.5\textwidth}
		\centering
		\includegraphics[width = 1 \textwidth]{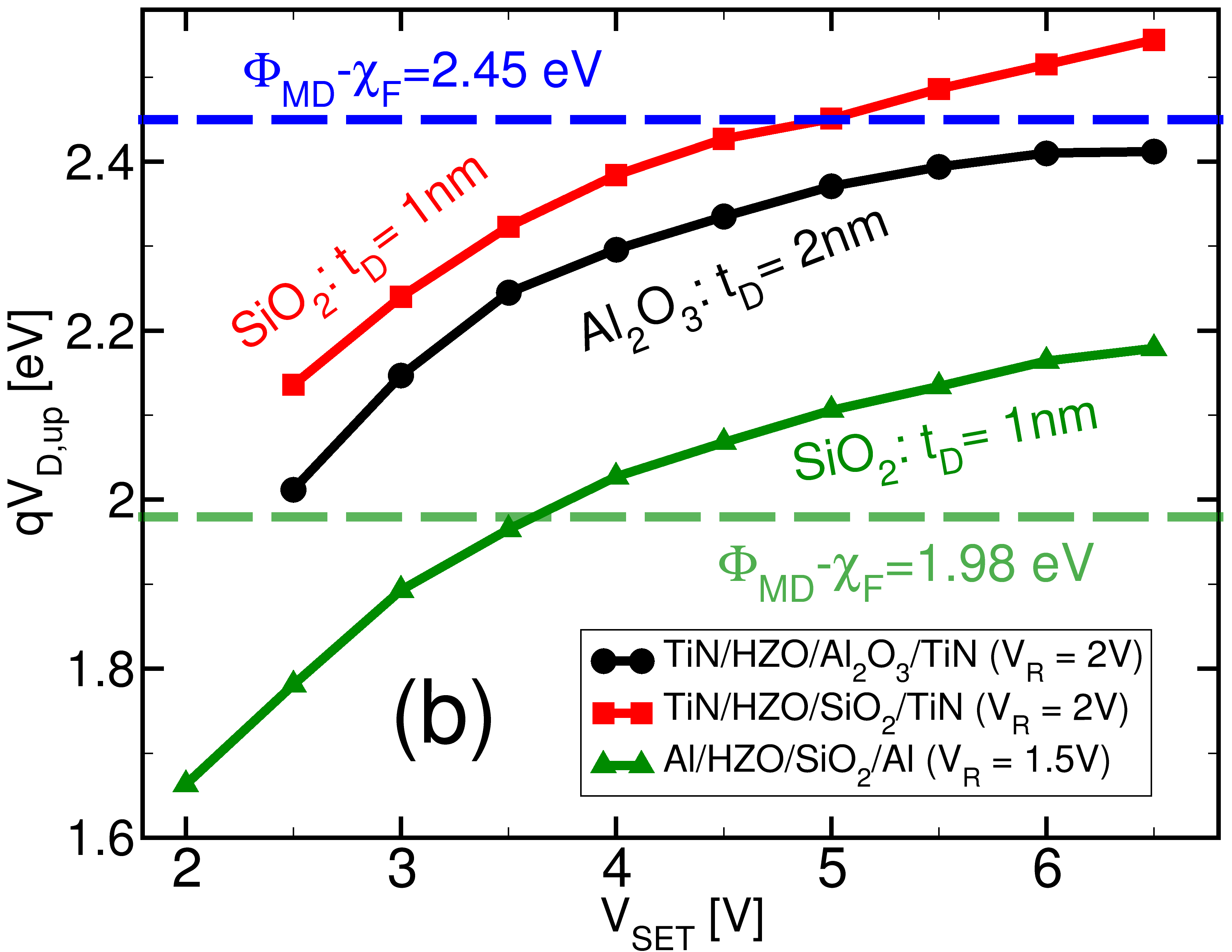}
	\end{minipage}
	\vspace{-2mm}	
	\caption{\label{Fig:Itunn_Options} \protect \footnotesize (a) Read current (left $y$-axis) and  $R_I$$=$$[I_{R,max}$$/$$I_{R,min}]$ (right $y$-axis) versus the set voltage for the TiN/HZO/Al$_2$O$_3$/TiN structure ($t_D$$=$2nm, $V_R$$=$2 V) in { Fig.\ref{Fig:Itunn_Exp_Sim}}, and for two variants of an HZO/SiO$_2$ based FTJ, namely for TiN/HZO/SiO$_2$/TiN ($t_D$$=$1nm, $V_R$$=$2 V), and for Al/HZO/SiO$_2$/Al ($t_D$$=$1nm, $V_R$$=$1.5 V). The HZO thickness is $t_F$$=12$nm in all cases. (b) The average voltage drop, $V_{D,up}$, for the positive polarization domains (as in { Fig.\ref{Fig:IT_Tox_Al2O3}}), for the different design options. The ferroelectric tunnelling barrier $[\Phi_{MD}$$-$$\chi_{F}]$ is substantially reduced for the Al electrode.
	}	
\end{figure}

In order to reduce $C_D$ for a given $t_D$ we explored SiO$_2$, having a dielectric constant about 2.5 times smaller than Al$_2$O$_3$. 
{Figure \ref{Fig:Itunn_Options}} reports $I_R$ and $R_I$ 
versus $V_{SET}$ for two variants of a HZO/SiO$_2$ based FTJ, and compared to the HZO/Al$_2$O$_3$ baseline case of { Fig.\ref{Fig:Itunn_Exp_Sim}} (black). By using a 1 nm SiO$_2$ layer and maintaining TiN electrodes with $V_{R}$$=$2.0~V (red) the $I_R$ largely increases, but $R_I$ does not improve. In the second option of the HZO/SiO$_2$ based FTJ we consider a low workfunction Al electrode ($\Phi_{M}$$\simeq$4.08 eV), so as to reduce the SiO$_2$ tunnelling barrier $[\Phi_{MD}$$-$$\chi_{D}]$. This leads to a large $I_R$ increase at fixed $V_R$, that we exploited to decrease $V_R$ to 1.5 V and the minimum $V_{SET}$ to 2 V. The corresponding results in {Fig.\ref{Fig:Itunn_Options}} (green) show a large improvement for both $I_R$ and $R_I$ compared to the HZO/Al$_2$O$_3$ case. Indeed, {Fig.\ref{Fig:Itunn_Options}}(b) reveals that the 1 nm SiO$_2$ design leads to $V_{D,up}$ values comparable to the 2 nm Al$_2$O$_3$ for $V_R$$=$2 V; moreover for Al electrodes the $qV_{D,up}$ can overcome $[\Phi_{MD}$$-$$\chi_{F}]$ even for $V_R$$=$1.5 V.
\section{Conclusions}
\label{sec:Conclusions}
By using an in-house developed simulator for FTJs calibrated against experiments, we studied the delicate tradeoffs between the reading efficiency and the depolarization effects for FTJs based synaptic devices. Our results show that a SiO$_2$ tunnelling dielectric and low workfunction metal electrodes can greatly increase the read current dynamic range and thus enable a multi-bit synaptic weight resolution.
\section*{Acknowledgement}
This work was supported by European Union through the BeFerroSynaptic project (GA:871737).
\clearpage

\input{references}
\clearpage
\onecolumn
\input{equations}

\input{table}

\vspace{-2mm}

\newpage

\end{document}

%% file: macros.tex
\newcommand{\kv}{{\bf k}}

\newcommand{\kx}{{\ensuremath{k_x}}}

\newcommand{\ky}{{\ensuremath{k_y}}}

\newcommand{\IR}{\ensuremath{\textrm{I}_{R}}}
\newcommand{\IRi}{\ensuremath{\textrm{I}_{R,i}}}



%% file: equations.tex
\begin{table}[t]
		\centering
		\begin{footnotesize}
			\renewcommand{\arraystretch}{2.2}
			\begin{tabular}[]{|p{18cm} p{4mm}|}
				\hline
			 	 $t_{F}\rho \dfrac{dP_i}{dt} = -(2 \alpha_i \, P_i + 4 \beta_i \, P_i^3+6 \gamma_i \, P_i^5)t_{F} - \dfrac{t_{F} \, k}{ d \, w} \sum_{n}(P_i - P_n ) - \dfrac{1}{2}\sum_{j=1}^{n_D}\left ( \dfrac{1}{C_{i,j}}+\dfrac{1}{C_{j,i}}\right ) \, P_j + \dfrac{C_D}{C_0} \, V_T$  & (1) \bigstrut \\[5pt]
				\hline
				$V_{D,i}$$=$$\dfrac{1}{d^2}\int_{D_i} V_D(\bar{r}) d\bar{r}$$=$$\sum_{j=1}^{n_D} \dfrac{1}{C_{i,j}} P_j$$+$$\dfrac{C_F}{C_0} V_T$, \hspace{5mm}  $V_{F,i}$$=$$V_T-V_{D,i}$$=$$ - \sum_{j=1}^{n_D} \dfrac{1}{C_{i,j}} P_j$$+$$\dfrac{C_D}{C_0} V_T$, 
				& (2) \\[5pt]
				\hline
			    $\IRi = \dfrac{q}{\pi \hbar}\int\displaylimits_{-\infty}^{\infty} \hspace{-1mm} \sum\limits_{k_x,k_y} \hspace{-1mm} T_i(E_{\perp}) [f_{0,MD}(E_{\perp} + \varepsilon(\kv)) - f_{0,MF}(E_{\perp} + \varepsilon(\kv))] \, dE_{\perp}$ \hspace{1mm} $\varepsilon(\kv) = (\hbar^2/2m_{\parallel})({k_x}^2+{k_y}^2)$
				& (3) \\[5pt]
				\hline
				$\sum\limits_{k_xk_y}f_{0,M} (E_\perp+\varepsilon(\kv)) = \dfrac{A}{(2\pi)^2} \int_{\kv} f_{0,M} (E_\perp+\varepsilon(\kv)) d\kv = \dfrac{A(K_B T)m_\parallel }{2\pi \hbar^2} \ln[1+\exp(\eta_{f,M})] $ \hspace{2mm} $\eta_{f,M}$$=$$\dfrac{(E_{f,M} - E_\perp )}{K_BT}$
				& (4) \\[5pt]
				\hline
				$\IRi = \dfrac{A(K_B T)m_\parallel q }{2\pi^2 \hbar^3} \int\displaylimits_{-\infty}^{+\infty} T_i(E_\perp) \left \{ \ln[1+\exp(\eta_{f,MD})]-\ln[1+\exp(\eta_{f,MF})] \right \} \, d E_\perp $
				& (5) \\[5pt]
				\hline
			\end{tabular}
		\end{footnotesize}
	\vspace{-5mm}
\end{table}

%% file: table.tex
\begin{table}[h]
	\centering
		%
		\begin{tabular}{|c|c|c|c|}
			\hline
			& $HZO$ & $Al_{2}O_{3}$  	& $SiO_{2}$   \\
			\hline
			\hline
			$\chi_{D}$, $\chi_{F}$ [eV]						&  	2.1	& 1.4 	& 0.95 	\\
			$m_{ox} [m_{0}]$								&   0.4 & 0.3	& 0.5  \\
			$\epsilon_{D}$, $ \epsilon_{F} [\epsilon_{0}]$	&  	30	& 10 	& 3.9 	\\
			\hline
		\end{tabular}
		\vspace{-0.15cm}\noindent\caption[]{\label{Tab:Sims_Param} \protect \footnotesize Material parameters used in simulations. The work function $\Phi_{M}$ of $Al$ and TiN were taken as $4.08 eV$ and $4.55 eV$, respectively.
		}
\end{table}